# Directing RF Terminals Using TELNET Applications

**Univ.Assist. Georgiana-Petruța Fîntîneanu**
**Faculty of Computers and Applied Computer Science**
**"Tibiscus" University of Timișoara, România**

**Rezumat.** Lucrarea de față își propune să evidențieze modul de utilizare și funcționare a protocolului TELNET pentru direcționarea terminalelor mobile. În acest sens lucrarea a fost structurată în trei părți: primele două părți au ca scop o scurtă prezentare teoretică a protocolului TELNET respectiv a terminalelor mobile, urmând ca în partea a III-a să fie prezentată o aplicație care evidențiază modul în care poate fi programat un terminal mobil folosind protocolul TELNET.

## 1. The TELNET protocol

Designed at the end of 1970 in the USA, the TELNET protocol allows the connection between a local terminal and a server in a network whit high connectivity and a wide variety of terminals.

It stands for TELEcommunications NETwork and it refers to both the application and protocol. Telnet offers the users a way of authentificating and accessing their terminals via the network. To put it more simply Telnet assures direct access to the remote computer. The access is made through port 23, port specifically reserved for the Telnet application but the Telnet server can be configured to use another port.

Telnet needs a Telnet server on the host computer which will be accessed. The Windows Operating Systems need a Telnet server in order to be accessed remotely. Computers whit UNIX O.S. or Linux derivative have a preinstalled Telnet Server.

An application whit text or graphic intelligence role is required on the Client Computer which connects through Telnet to a server. The Telnet application offered by the operating system is called Telnet. Windows and Linux alike offer a simple application in text mode which permits connection through Telnet.

The Telnet Protocol is a TCP (Transmission Control Protocol) connection used for transmitting data, containing Telnet control information.

The Telnet Protocol allows the user to identify himself on a remote system through the local system. This protocol establishes a client-server relationship between the local system (client) and the remote Telnet application (server), permitting the functioning of a local system as a virtual terminal connected to a virtual system. The Telnet Protocol serves for simulating a terminal connected to a Computer System.

The real terminal is shown in the image bellow:

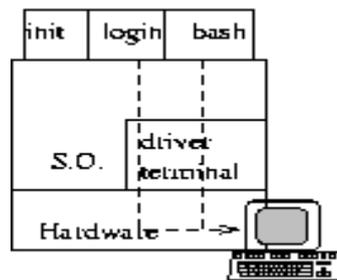

*Figure 1. The real terminal*

The Telnet mechanism simulates on a computer in a network, a terminal connected to the Telnet server.

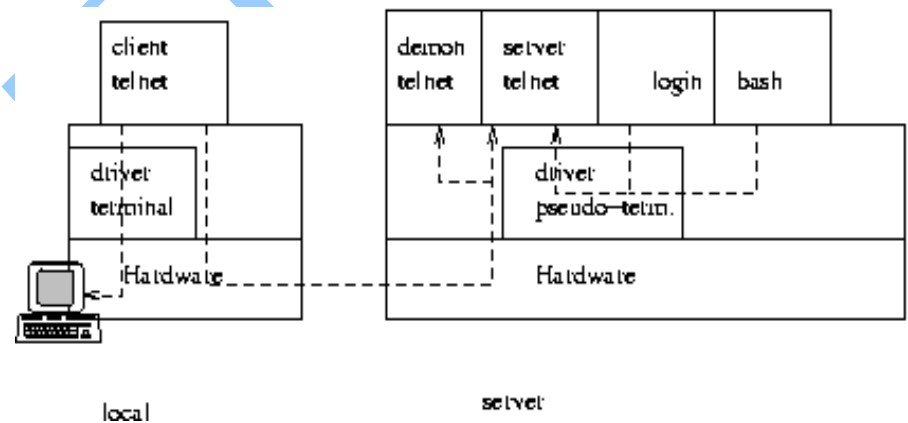

*Figure 2. The telnet mechanism*

## 2. RF terminals (Radio Frequency)

The RF system represents a wireless data via digital radio signals, at a certain frequency. The RF establishes a real time, bidirectional, radio connection between a, mobile terminal and the host computer. The mobile portable terminal can be manoeuvred by a worker or mounted in a forklift. It collects and shows data from the place where the work is being carried out. The host computer can be a PC, a minicomputer or even a more complex server. The end result is a flux of information to and from the host computer, allowing the worker to go to wherever he is needed without worrying about the fact that he may not have the data he needs to complete his task.

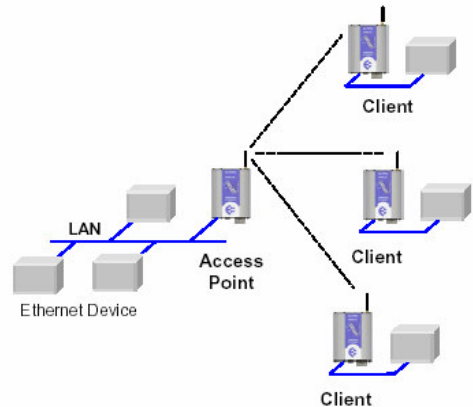

*Figure 3. The network between RF terminals and server*

The mobile terminal acts as an interface between the user and the RF system. It collects the data for transmission, receives instructions or data from the host computer and permits the user to view the data on the display.

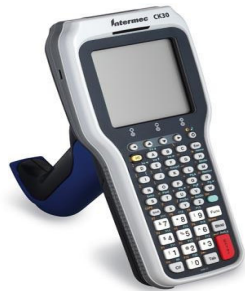

*Figure 4. The RF terminal*

63

The terminal also has a bidirectional antenna to ensure communication whit the rest of the system.

The base station has an antenna system which acts as a link between the wireless networks and fixed ones. It's connected to a controller, (the controller con be separate or integrated in the base station), which itself is connected to a host computer. The controller receives and processes the information from the host computer and transmits the information to the terminal through the base station.

The Telnet protocol offers users a way of authentication an accessing their remote terminals via the network. To put it more simply, the user accesses the remote computer directly through the wireless network using a curtain interface.

This type of reader requires the existence of a mobile phone type network. A central server has one or mare RF nodes (transceivers) attached; the number of nodes required depends on the on the area that needs to be covered and the building's properties which can affect RF transmissions. One or more mobile RF terminals communicate continuously with the server. Generally these types of terminals have a keyboard and display to permit the operator to obtain and transmit from ant to the computer data in a facile form. On these types of equipments you can install dedicated applications or you have the option every terminal acting as a workstation emulating parts or the application on the computer.

This last type of interface can be used when we wish to communicate whit a remote computer through the Telnet protocol.

## 3. Applying TELNET in development of RF mobile terminals user interface

We will present the UIM application which highlights the way a mobile terminal can be programmed using the Telnet protocol.

UIM stand for User Interface Mobile and is inspired from the purpose of the package, easing the design of user interfaces for creating applications dedicated to using RF mobile terminals.

Application from the logistic and storing domains are complex systems which are comprised of both hardware and software components.

**UIM hardware architecture**
The hardware components are represented in the figure below:

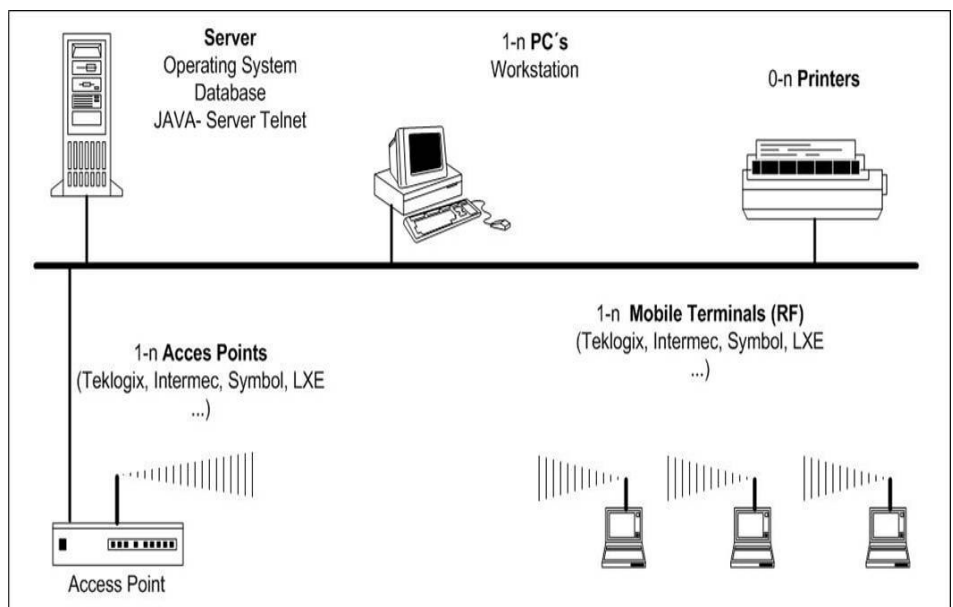

*Figure 5. UIM hardware architecture*

From a hardware point of view, the system is made up of an application server, the network, workstations, printers and mobile terminals.

Installed on the server is the most important software component of the system. For the user interface part, the server has installed the UIM application and a storage medium of the fluxes and screens (a database).

The workstations are used to administer the system and monitoring the activity of the mobile terminal users. On the workstation level are software applications specific for communication whit the server.

Based on the specifics of the applications, printers can be used for document and reports. The mobile RF terminals are the ones responsible to guide the personals activity and transmit data to the server. The server in linked to the printers and workstations via the fixed network and via wireless network to the mobile terminals (represented by access points). The link between the mobile terminals and the rest of the equipment is made trough the server and the UIM application (the Telnet server). From the software architecture point of view the present work presents only the UIM application necessary for the interface of the RF mobile terminals.

### UIM Software Architecture
UIM offers a series of software libraries and implementation forms that a "beginner" programmer can use to quickly design user interfaces to exploit mobile terminals. So, for designing interfaces you can just create some XML files or you can opt to configure some interfaces using tables in a database and the XML files will be generated automatically based on the saved data. UIM takes the data in XML format and generates the interface based on the data.

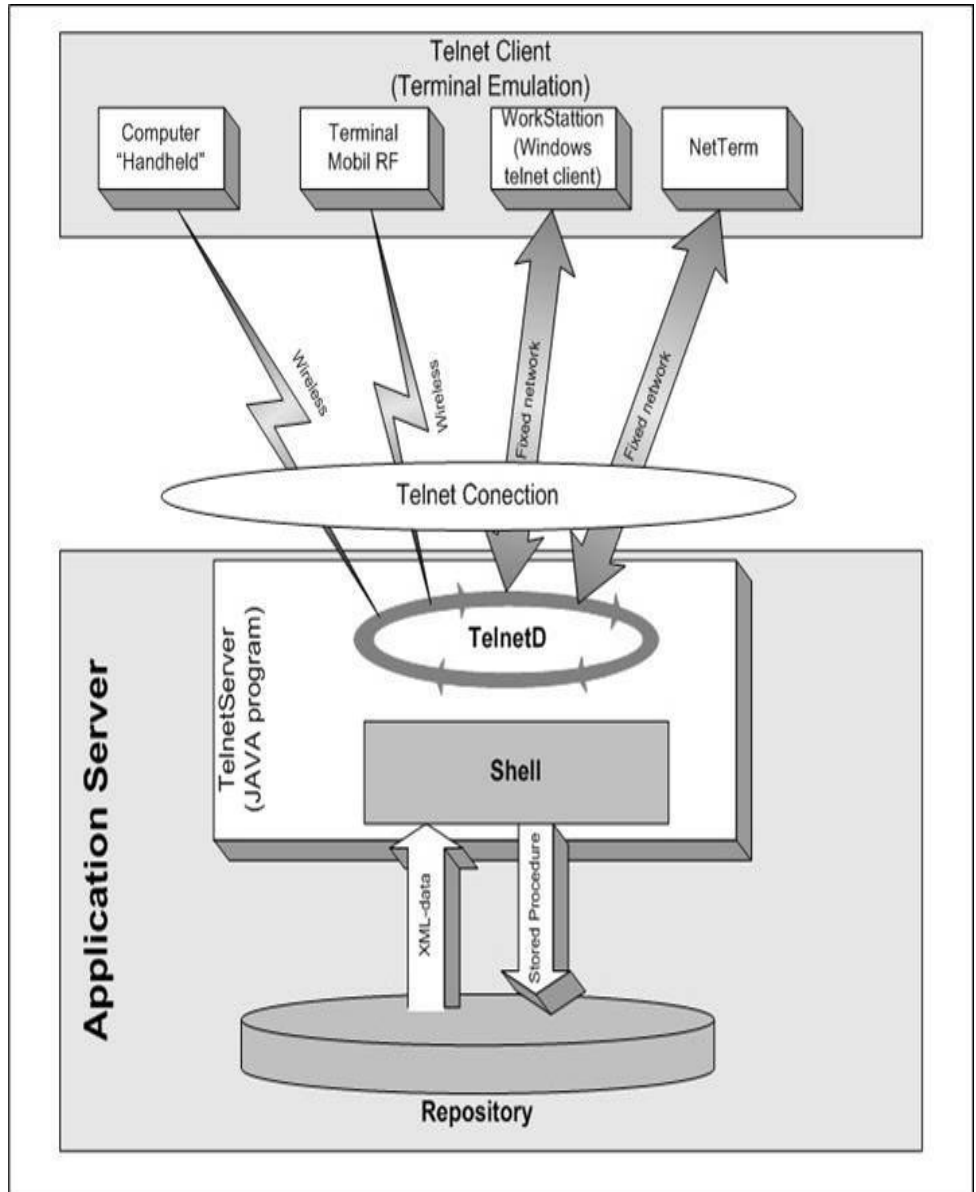

*Figure 6. Software architecture*

The main component of the architecture is the Telnet Server witch is developed based on an "Open Source" library called TelnetD. When a client is making a request a TelnetD session is created by TelnetD software witch implements the Telnet protocol at a general level. Being an "Open Source" product it could be modified to call a class that give the control to the Shell.

The Shell verifies the repository and took the information depending on what was the client request and displays this info on the mobile terminals (the terminals can be computer "Handheld", mobile terminal RF, Workstation (Windows telnet client), NetTerm). The connection between the terminal and the server can be done throw a network wireless or wired. The repository can be a database or a system file and have to return dates in XML format.

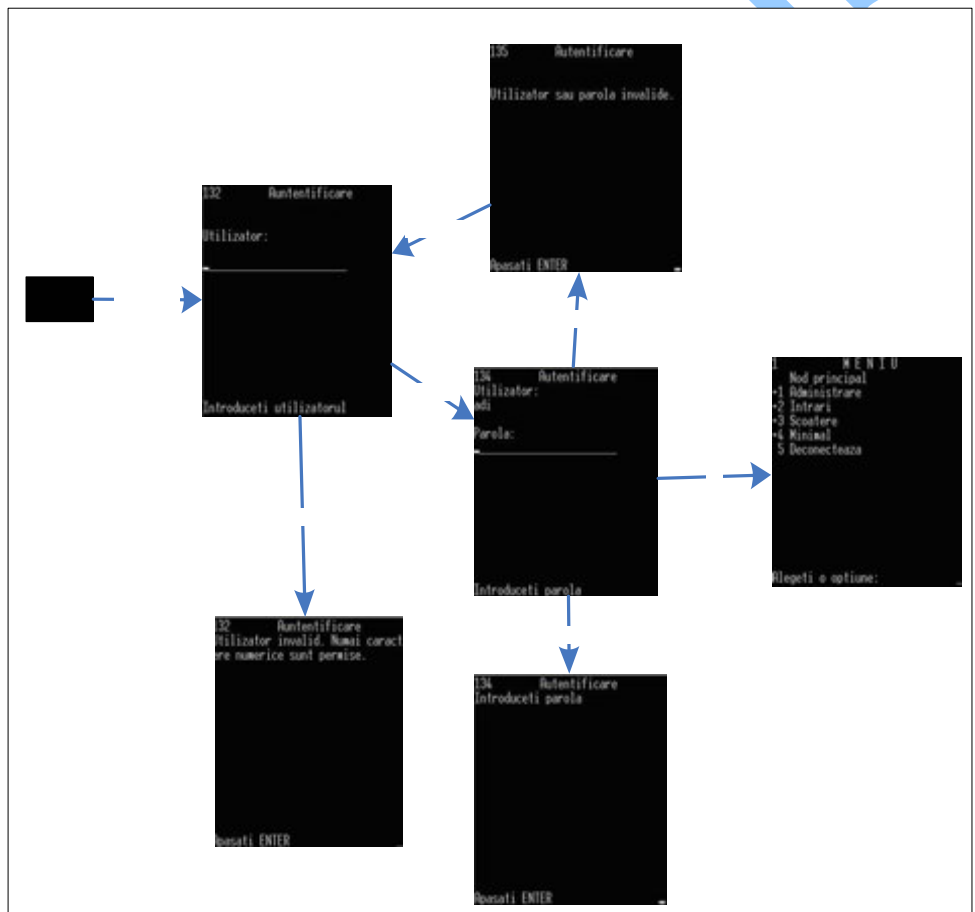

*Figure 7. Flow*

These RF terminals are used by simple workers, so the interface is made as simpler as possible.

We made five types of screens: Menu Screen, Info Screen, Input Screen, Single-option Screen, Multi-options Screen. The main screen is Menu Screen that has a tree organisation with leaves and nodes. Every line of menu represents an option. When the line started with plus meaning that is a nod, else is a leave. For selecting one operation they have to type the number before that line. For coming back they have to type zero. Info Screen is used to display information. With the Input Screen users have the possibility to insert information in applications.

Depending of the task they need the user can create a flow with this screens.

Keeping up to date and collecting date in real time is critical for organizing and manipulating data in warehouses and stores. The application presented in this article offers a way to easily implement and use systems that allows mobility and accuracy.